\documentclass[11pt]{article}
\usepackage[centertags]{amsmath}
\usepackage{amsfonts} \usepackage{amssymb} \usepackage{amsthm}
\def\one{{\rm 1\kern -.9mm l}}                             %
\newcommand{\braket}[2]{\langle{#1}|{#2}\rangle}           %
\usepackage{graphicx}
\usepackage{dcolumn}
\newcommand{\ov}{\overline}
\begin{document}
\begin{titlepage}
\rightline{DSF-19/2002} \rightline{NORDITA-2002/48 HE} \vskip
3.0cm \centerline{\LARGE \bf $\mathbf{\mathcal{N}}=1$  Matter from
Fractional Branes  } \vskip 1.4cm \centerline{\bf  R. Marotta
$^a$, F. Nicodemi $^a$,  R. Pettorino $^a$, F. Pezzella $^a$ and
F. Sannino $^b$} \vskip .8cm \centerline{\sl $^a$ Dipartimento di
Scienze Fisiche, Universit\`a di Napoli and INFN, Sezione di
Napoli} \centerline{\sl Complesso Universitario Monte
S. Angelo - via Cintia -  I-80126 Napoli, Italy} \vskip .4cm \centerline{\sl $^b$
NORDITA, Blegdamsvej 17, DK-2100 Copenhagen \O, Denmark} \vskip
2cm

\begin{abstract}
We study a bound state of fractional D3-branes localized inside
the world-volume of fractional D7-branes on the orbifold
$\mathbb{C}^{3} / ( \mathbb{Z}_{2} \times \mathbb{Z}_{2} )$. We
determine the open string spectrum that leads to ${\cal N}=1$
$U(N_1) \times U(N_2) \times U(N_3) \times U(N_4)$ gauge theory
with matter having the number of D7-branes as a flavor index. We
derive the linearized boundary action of the D7-brane on this
orbifold using the boundary state formalism and we discuss the
tadpole cancellation. After computing the asymptotic expression of
the supergravity solution the anomalies of the gauge theory are
reproduced.
\end{abstract}
\vfill  {\small Work partially supported by the European
Commission RTN Programme HPRN-CT-2000-00131 and by MIUR.}
\end{titlepage}
\newpage
\tableofcontents       %
\vskip 1cm             %

\section{Introduction}
\label{introduction} An efficient way to study the relation
between String Theory and Standard Model (SM) field theory is
through the {\it bottom-up} approach \cite{AIQU} which constitutes
an alternative to the traditional {\it top-down} approach. In the
latter one starts from a string theory and tries to reproduce the
SM massless spectrum by a suitable Calabi-Yau manifold compactification that reduces the number of space-time dimensions and supersymmetry, giving the right gauge group. In the former,
instead, one looks for configurations of D-branes with
world-volume gauge theories similar as much as possible to the SM.
One of the advantages of this approach is that it does not require
any knowledge of the geometrical details of the manifold
compactification. We would like here to observe that the bottom-up
approach is essentially based on the property that a D-brane
supports a gauge field theory on its world-volume. On the other
hand it is well-known that D-branes can be studied from a
complementary point of view: they are classical solutions of the
ten-dimensional supergravity. This is the essence of the {\it
gauge/gravity} correspondence. It means that one can analyze the
classical backgrounds generated by a particular configuration of
D-branes to get some insight on the {\it dual} gauge theories
living on their world-volume and viceversa. This is just the point
of view we want to adopt in this paper that can represent a sort
of link between the bottom-up approach and the gauge/gravity
correspondence. We study the latter by considering a suitable
configuration of fractional branes yielding the SM-like gauge
theory. Fractional branes \footnote{For a recent review on
fractional branes  see Ref.~\cite{Bertolini:2001gq} and references
therein.}, living at orbifold singularities, are interesting tools
to study gauge/gravity correspondence for non conformal gauge
theories with reduced supersymmetry
\cite{{Bertolini:2001gq},Klebanov:1999rd} in different
renormalization schemes \cite{Marotta:2002ns}. This aim would be
better achieved, for ${\cal N}=1$ theories, by considering branes
wrapped on supersymmetric cycles inside a Calabi-Yau space. In
fact it has been shown that the classical supergravity solutions
determined by these non perturbative objects are completely
regular \cite{KS,MN} and therefore they allow to study also the
infrared properties of the dual gauge theories
\cite{{ABCPZ},DLM,I,OS}. Unfortunately, within this approach, it
is not straightforward to find a classical background dual to
gauge theory with chiral matter, which is instead more easily
deducible from bound states of fractional branes. We expect, in
analogy with similar configurations, that the classical background
determined by fractional branes
\cite{Bertolini:2000dk,Polchinski:2000mx,Frau:2000gk,Billo:2001vg,Grana:2000jj,Bertolini:2001qa}
 contains a naked singularity resolved by the
{\it enhan\c{c}on} mechanism \cite{{Johnson:1999qt}}. The
enhan\c{c}on is a sort of screen which excises the region of
space-time corresponding to scales where non-perturbative effects
become relevant in the gauge theory  and this clarifies why we are
able to obtain only the ultraviolet behavior of the field theory.

In this paper we consider a stack of fractional D3-D7 branes
living on the orbifold $\mathbb{C}^{3} / ( \mathbb{Z}_{2} \times
\mathbb{Z}_{2} )$. The choice of such an orbifold is motivated by
the interest in ${\cal N}=1$ supersymmetric gauge theories with
matter. In fact this is one of the simplest orbifolds preserving
four supersymmetry charges. {}Furthermore in this orbifold
background there are four different kinds of fractional D-branes
corresponding to the four irreducible representations of the
abelian discrete group $\mathbb{Z}_{2} \times \mathbb{Z}_{2}$
\cite{{Diaconescu:1999dt},{BDFM}}. The gauge theory supported by a
general bound state of $N_{I}$ fractional branes of type $I$ is a
$\bigotimes_{I=1}^{4} U(N_{I})$ gauge theory with chiral
multiplets charged under the bifundamental representations of the
gauge group. The inclusion of fractional D7-branes, since the open
strings have one end on a D3-brane and the other end on a D7,
allows us to have chiral multiplets charged only under the
fundamental representation of the gauge group. They have also a
flavor index running on the number of fractional D7-branes.

It is relevant to note that when we turn off three of the gauge
groups by choosing $N_2=N_3=N_4=0$ the gauge theory (in the
presence of the D7-brane) is exactly ${\cal N}=1$ super QCD with
matter.

In the forthcoming discussion we will consider a bound state of a
D3-brane living at the orbifold singularities and hence it is
completely localized in the world-volume of the D7-brane. The
latter is partially extended along the orbifold directions. {}For
our purposes it has been enough to compute the asymptotic behavior
of the classical supergravity solution via the boundary state
formalism \cite{DiVecchia:1999rh,{DiVecchia:1999fx}}.

The solution so found exhibits a $U(1)$ isometry acting as a phase
factor on the coordinates transverse to the D3 branes. Such an
invariance is related to the classical $\hat{U}(1)_{R}$ R-symmetry
of the dual gauge theory \cite{BDFLM02}. In a gauge theory this
symmetry is just a member of a supersymmetric multiplet containing
also the dilations. In classical theories, which are invariant
under superconformal transformations, the corresponding currents
are conserved. Quantum corrections break superconformal invariance
and give rise to anomalies. The anomaly breaking of the global
$\hat{U}{(1)}_{R}$ symmetry appears as a spontaneous symmetry
breaking in supergravity \cite{Klebanov:2002gr}. It is important
to stress that the $\hat{U}(1)_R$ is not the final anomaly free
$U(1)_R$ which is obtained via a suitable choice of the R-charges
of the various fields. We have found that also our classical
solution explicitly breaks the $U(1)$ isometry to one of its
discrete subgroups. Therefore we expect to obtain the associated
anomaly of the underlying gauge theory. In fact, by analyzing on
the supergravity side how the solution transforms under a combined
scale and chiral transformation, we have been able to reproduce
the corresponding anomalies, i.e. the one-loop $\beta$-function
and its chiral anomaly. It is evident that we have used
supergravity to get perturbative information on the dual gauge
theory. This means that our correspondence has to be interpreted
not in the same spirit as AdS/CFT, but more properly in that of
open/closed string duality.

The paper is organized as follows.

In sect. 2 we analyze the properties of fractional D3-branes on
the orbifold $\mathbb{C}^3 /(\mathbb{Z}_2  \times \mathbb{Z}_2) $
and in particular we study how the orbifold acts on the open
strings stretched between two of these branes. This analysis
allows us to obtain the massless spectrum of the gauge theory
surviving the orbifold projection.

{}Furthermore we introduce fractional D7-branes, determine the massless
spectrum of the matter chiral fields and fix a configuration of D7-branes
that leads to an anomaly free gauge theory.

In sect. 3 we analyze, from the point of view of the boundary
state approach, the boundary action of our system. In particular,
by writing the boundary states of a fractional D7-brane and by
computing the coupling of this boundary with the bulk fields, we
construct, at linearized level, the boundary action for  such a
brane in this orbifold. {}Furthermore, still within the boundary
state formalism, we also compute the classical solution of the
equations of motion at first order in the string coupling $g_s$.
In the last part of this section we investigate the stability of
the system related to the one-loop massless tadpole cancellation.

{}Finally in sect. 4, by studying how the scale and chiral transformations of
the dual gauge theory are realized in supergravity, we are able to compute
the corresponding anomalies. The one-loop $\beta$-function and the chiral
anomalies emerge out from the asymptotic classical profile of the bulk fields.

\section{D3 and D7-branes  on the orbifold $\mathbb{R}^{1,3} \times
\mathbb{C}^3
/(\mathbb{Z}_2  \times \mathbb{Z}_2) $ }

Let us consider a system of D3 and D7-branes of type IIB string theory on the orbifold
$R^{1,3} \times \mathbb{C}^3 / (\mathbb{Z}_2 \times \mathbb{Z}_2) $.
In particular we analyze
the spectrum of the massless open string states having their end-points attached
to them. We take the orbifold directions to be along $x^4, \dots, x^9$.

\subsection{ General features of the orbifold
$ \mathbb{C}^3 / ( \mathbb{Z}_2 \times \mathbb{Z}_2 ) $ }

The six coordinates of the orbifold space can be simply arranged in three
complex coordinates spanning $\mathbb{C}^3$:
\begin{eqnarray}
z^1=x^4+ix^5, \quad z^2=x^6+ix^7, \quad z^3=x^8+ix^9 \ .
\end{eqnarray}
The $\mathbb{Z}_2$ group is characterized by two elements
$\{\mathbf{1}, h \}$,  with $h^2=\mathbf{1}$, hence the
tensor product $\mathbb{Z}_2\times \mathbb{Z}_2$ is made of the
following four elements
\begin{eqnarray}
 \mathbb{Z}_2\times
\mathbb{Z}_2 \, : \, \{e=\mathbf{1}\times \mathbf{1}, \quad   h_1=
h\times \mathbf{1}, \quad h_2=\mathbf{1}\times h, \quad
h_3=h\times h\} \ . \label{group}
\end{eqnarray}
The  action of the orbifold group $\mathbb{Z}_2 \times \mathbb{Z}_2$
on the complex vector $\vec{z}=(z^1,z^2,z^3)$
of $\mathbb{C}^3$
is given by a three-dimensional representation (necessarily
reducible) of this finite group.
We perform the following choice:
\begin{eqnarray}
R_e= \left[
\begin{array}{ccc}
~~1 & ~0& ~~0
\\ ~~0& ~1 & ~~0 \\ ~~0 & ~0  & ~~1
\end{array}
\right],  &&
\quad  R_{h_1}=\left[
\begin{array}{ccc}
1 & 0& 0
\\ 0& -1 & 0 \\ 0 & 0  & -1
\end{array}
\right], \nonumber\\
 \quad R_{h_2}= \left[
\begin{array}{ccc}
-1 & 0& 0
\\ 0& 1 & 0 \\ 0 & 0  & -1
\end{array}
\right],
&&\quad   R_{h_3}= \left[
\begin{array}{ccc}
-1 & 0& 0
\\ 0& -1 & 0 \\ 0 & 0  & 1
\end{array}
\right] \ .
\end{eqnarray}
It is possible to summarize the orbifold action on $\mathbb{C}^3$
as follows:
\begin{eqnarray}
R_{h_i} = {\rm diag} \left[ e^{i\pi b^1_{h_{i}}},e^{i\pi
b^2_{h_{i}}},e^{i\pi b^3_{h_{i}}} \right] \ ,
\end{eqnarray} with
\begin{eqnarray}
\vec{b}_{h_1}=(0,1,1),\quad \vec{b}_{h_2}=(1,0,1),\quad
\vec{b}_{h_3}=(1,1,0) \ . \label{bhi}
\end{eqnarray}
The vectors defined in (\ref{bhi}) clearly satisfy the following
condition: \begin{equation} b^1_{h_i} + b^2_{h_i} + b^3_{h_i} = 0
\qquad {\rm mod.} 2 \ .
\end{equation} The present choice of the orbifold representation from the
point of view of the gauge theory is the one which leaves intact
the ${\cal N}=1$ theory we are interested in.

The spectrum of the closed string theory on the orbifold space
$\mathbb{C}^{3}/(\mathbb{Z}_{2} \times \mathbb{Z}_{2})$ consists
of an untwisted sector, corresponding to the identity of the
orbifold group and three twisted sectors corresponding to its non
trivial elements. In particular, the massless spectrum in the
twisted sectors coincides with the zero modes of supergravity
fields dimensionally reduced on the three exceptional vanishing
two-cycles ${\cal C}_{i}$ with $i=1,2,3$ characterizing the
orbifold, each of them embedded in one of the three
four-dimensional subspaces of $\mathbb{C}^{3}$. The three anti
self-dual two-forms $\omega_{2}^{i}$, dual to the cycles ${\cal
C}_{i}$, are completely independent and normalized as:
\footnote{In this paper we adopt a different convention with
respect to the paper of Ref.~\cite{BDFM}. One can go from one
convention to the other by rescaling by a factor 2 the
antiself-dual two-forms: $\omega = 2\omega_{ \mbox{\cite{BDFM}} }
$.}
\begin{equation}
\int_{{\cal C}_{i}} \omega_{2}^{j} = 2 \delta^{i}_{j}, \, \, \,
\int \omega^{i}_{2}
\wedge \omega^{i}_{2} = -1, \, \, \,  *_{4} \omega^{i}_{2} = - \omega^{i}_{2}
\end{equation}
where $*_{4}$ denotes the dual in the four-dimensional space in
which the two-cycle is embedded.

In the following we are interested in describing fractional branes coupled
to two kinds of twisted scalars and twisted RR four-form. The scalars
are obtained by
reducing the Kalb-Ramond two-form
$B_{2}$ and the RR two-form on the exceptional two-cycles, while the RR
four-form is obtained by reducing the RR six-form, i.e.:
\begin{equation}
B_{2} =  \sum_{i=1}^{3} b^{i} \omega_{2}^{i};\, \, \,
C_{2}=\sum_{i=1}^{3} c^{i} \omega_{2}^{i} \,; \, \, C_{6} =
\sum_{i=1}^{3} A_{4}^{i} \wedge \omega_{2}^{i} \ .
\end{equation}

\subsection{ D3-D3 open string spectrum on the orbifold $\mathbb{C}^{3}/(\mathbb{Z}_{2}
\times \mathbb{Z}_{2}) $ }

Let us introduce now fractional D3-branes on this orbifold. In particular we
are interested in studying configurations where these branes are transverse
to the orbifold, namely with world-volume directions $x^{\mu}$ with
$\mu=0,1,2,3$.

Before the orbifold projection, the low energy modes consistent
with the open superstring theory living on the $N$ $D3$-branes,
are those of the four dimensional ${\cal N}=4$ $U(N)$ super Yang
Mills theory. In fact a generic open string state is the product
of a Chan-Paton factor, that is in this case an $N \times N$
matrix, and of an oscillator part. In particular, in the
Neveu-Schwarz sector, the massless states are given by:
\begin{eqnarray}
&&A^{\mu}= \lambda \otimes \psi_{-1/2}^{\mu}|0, k \rangle, \nonumber\\
&&\phi^i= \lambda \otimes \left(\psi_{-1/2}^{2i+2}+ i \psi_{-1/2}^{2i+3}\right)
|0,k \rangle
\label{nsst}
\end{eqnarray}
where $\lambda$ denotes the Chan-Paton factor and $i=1,2,3$.

The states $A^{\mu}$ can be identified with the $U(N)$ gauge bosons of
the supersymmetric ${\cal N}=4$ gauge theory; the three complex fields
$\phi^i$ describe the six real scalars living, as $A^{\mu}$, in the
adjoint representation of the $U(N)$ gauge group. All of them fill the
bosonic part of the ${\cal N}=4$ gauge multiplet.

The same analysis can be repeated for the Ramond sector. In this case the
massless spectrum is identified with the four
two-component (on shell) fermions still in the adjoint representation
of $U(N)$.

Let us now consider the spectrum in the orbifold theory, taking
into account that the orbifold group acts both on the oscillators
and on the Chan-Paton factors. The only allowed states are the
ones surviving under this combined action. We would like here to
remind that the action of the orbifold group element $h$ on the
Chan-Paton factors is defined as:
\begin{equation}
\gamma(h) \lambda \gamma^{-1}(h) = \lambda^{'}.
\end{equation}
In the case of $\mathbb{Z}_{2} \times \mathbb{Z}_{2}$, $h$ runs
over the four elements of this group defined in (\ref{group}).

We are interested in {\it fractional} branes which are defined as
D-branes whose Chan-Paton factors transform under the irreducible
representations of the orbifold group. $\mathbb{Z}_{2} \times
\mathbb{Z}_{2}$ has four one-dimensional irreducible
representations:
\begin{equation}
\begin{array}{llll}
\gamma_{1} (e)= + 1 & \gamma_{1}(h_{1})=+1 & \gamma_{1}(h_{2})=+1 &
\gamma_{1} (h_{3}) = + 1, \\
\gamma_{2} (e)= + 1 & \gamma_{2}(h_{1})=+1 & \gamma_{2}(h_{2})=-1 &
\gamma_{2} (h_{3}) = - 1, \\
\gamma_{3} (e)= + 1 & \gamma_{3}(h_{1})=-1 & \gamma_{3}(h_{2})=+1 &
\gamma_{3} (h_{3}) = - 1, \\
\gamma_{4} (e)= + 1 & \gamma_{4}(h_{1})=-1 & \gamma_{4}(h_{2})=-1 &
\gamma_{4} (h_{3}) = + 1. \\
\end{array} \label{action}
\end{equation}
The above definition implies that fractional branes, differently
from the regular ones, do not admit images and hence they live at
the orbifold fixed point $z_{1}=z_{2}=z_{3}=0$. {}Furthermore it
also shows that there are four different kinds of fractional
branes that we label with the index $I=1,2,3,4$. In the following
we consider a configuration made of $N_{I}$ D3-branes of type $I$,
with $\sum_{I=1}^4 ~N_{I}~=N$. The orbifold action in
(\ref{action}) is generalized to:
\begin{eqnarray}
\gamma^{(D3)}_{e}&=&{\rm diag}(\mathbf{1}_{N_1, N_1},\,\,\,\,
\mathbf{1}_{N_2, N_2},\,\,\,\, \mathbf{1}_{N_3, N_3},\,\,\, \,\mathbf{1}_{N_4, N_4} ), \\
\gamma^{(D3)}_{h_1}&=& {\rm diag}(\mathbf{1}_{N_1, N_1},\,\,\,\,
\mathbf{1}_{N_2,N_2},-\mathbf{1}_{N_3,N_3}, - \mathbf{1}_{N_4,N_4} ), \\
\gamma^{(D3)}_{h_2}&=&{\rm diag}(\mathbf{1}_{N_1,N_1},-\mathbf{1}_
{N_2, N_2},\,\,\,\,\mathbf{1}_{N_3,N_3}, -\mathbf{1}_{N_4,N_4}), \\
\gamma^{(D3)}_{h_3}&=&{\rm diag}(\mathbf{1}_{N_1,
N_1},-\mathbf{1}_{N_2,
N_2},-\mathbf{1}_{N_3,N_3},\,\,\,\,\mathbf{1}_{N_4,N_4})  .
\end{eqnarray}
Our orbifold action preserves only the ${\cal N}=1$ supersymmetry
while acting on the states of the NS-sector as follows:
\begin{eqnarray}
&&\left[ \gamma^{(D3)}_{h_l} \right] A^{\mu} \left[ {\gamma^{(D3)}_{h_l}} \right]^{-1}=
A^{\mu} ,
\\
&&\left[ \gamma^{(D3)}_{h_l} \right] \phi^i\, \,
\left[ {\gamma^{(D3)}_{h_l}} \right]^{-1}R_{h_l}^{ij}=  \phi^j .
\label{NS}
\end{eqnarray}
The physical states invariant under the orbifold action and which can be
identified as gauge bosons are the following:
\begin{equation}
A_\mu =
\begin{pmatrix}
A_\mu^1 & 0        & 0       & 0        \cr  0        & A_\mu^2
& 0       & 0        \cr  0        & 0        &  A_\mu^3 & 0 \cr  0 &
0        &  0      & A_\mu^4  \cr
\end{pmatrix}
\end{equation}
showing that the gauge group is now $\displaystyle{U(N_1)\times
U(N_2)\times U(N_3) \times U(N_4)}$, while the scalars of the
twelve chiral multiplets can be organized in the following
matrices:
\[ \phi^1 = \begin{pmatrix}0 & a_{ i_1, \ov{j}_2} & 0   & 0   \cr
a_{i_{2}, \ov{j}_1} & 0 & 0   & 0 \cr
                    0 & 0   & 0   & a_{i_{3}, \ov{j}_4} \cr  0   & 0
& a_{i_{4}, \ov{j}_3} & 0
                    \cr\end{pmatrix}\;,\; \phi^2 = \begin{pmatrix}0 &
                    0   & a_{i_{1}, \ov{j}_3} & 0  \cr     0   & 0   & 0
& a_{i_{2}, \ov{j}_4}\cr
                    a_{i_{1}, \ov{j}_3} & 0   & 0   & 0  \cr  0
& a_{i_{4}, \ov{j}_2} & 0 & 0
                    \cr\end{pmatrix} \]
\begin{equation}
\phi^3 = \begin{pmatrix}0 &
                    0   & 0   & a_{i_{1}, \ov{j}_4} \cr  0   & 0
& a_{i_{2}, \ov{j}_3} & 0   \cr  0 &
                    a_{i_{3}, \ov{j}_2} & 0   & 0
\cr  a_{i_{1}, \ov{j}_4} & 0   & 0   & 0
                    \cr\end{pmatrix}
\end{equation}
where $i_{I}=1, \dots, N_{I}$ and the bar on the index $j_{I}$
picks in the antifundamental representation of the group. Since we
preserve ${\cal N}=1$ it is sufficient to consider the orbifold
action on the NS sector and to deduce via supersymmetry the
fermionic spectrum in the R sector.

In Table \ref{tavola1} we summarize the spectrum of the chiral
superfields according to their transformation properties under
each gauge group.
\begin{table}
\begin{eqnarray*}\begin{array}{cccc}
  U(N_1) & U(N_2) & U(N_3) & U(N_4) \\
  \hline \hline \\
  \square & \overline{\square}& 1& 1 \\
  \overline{\square}& \square & 1 & 1 \\
  \square& 1&\overline{\square} & 1   \\
  \overline{\square}&1 & \square & 1  \\
  {\square} & 1&1& \overline{\square} \\
  \overline{\square} & 1&1& {\square} \\
    1&{\square}& \overline{\square}&1 \\
 1&\overline{\square}& {\square}&1 \\
 1&{\square}&1& \overline{\square} \\
 1&\overline{\square}&1& {\square} \\
  1&1&{\square}& \overline{\square} \\
 1&1&\overline{\square}& {\square} \\
\end{array}
\end{eqnarray*}
\caption{Spectrum of the chiral superfields. $\square$ and
$\overline{\square}$ denotes respectively the fundamental and the
antifundamental of each gauge group.} \label{tavola1}
\end{table}
It can be written compactly:
\begin{eqnarray}
\Phi^1 &\equiv& (N_1,\overline{N}_2) +
(N_2,\overline{N}_1)+(N_3,\overline{N}_4) + (N_4,\overline{N}_3) \ , \\
\Phi^2 &\equiv & (N_1,\overline{N}_3) +
(N_2,\overline{N}_4)+(N_3,\overline{N}_1) + (N_4,\overline{N}_2) \ , \\
\Phi^3 &\equiv& (N_1,\overline{N}_4) +
(N_2,\overline{N}_3)+(N_3,\overline{N}_2) + (N_4,\overline{N}_1) \ .
\end{eqnarray}
As expected, we have a vector-like theory which does not suffer of
any gauge anomaly and it is well-defined. The vectors together
with the relative gauginos (R-sector) are all living in the adjoint
representation of each group. The corresponding one-loop
$\beta$-functions  are:
\begin{eqnarray}
\beta{(g_1)} &=&-\frac{g^3_1}{16\pi^2} (3N_1-N_2-N_3-N_4)\ , \\
 \beta{(g_2)}&=& -\frac{g^3_2}{16\pi^2}(3N_2 -N_1-N_3-N_4)\ ,   \\
 \beta{(g_3)}&=& -\frac{g^3_3}{16\pi^2}(3N_3 - N_1-N_2-N_4)\ ,     \\
 \beta{(g_4)}&=&-\frac{g^3_4}{16\pi^2} (3N_4 -N_1-N_2-N_3) \ .
\end{eqnarray}
It is worth noticing that all of them vanish when
$N_1=N_2=N_3=N_4$.

\subsection{D3-D7 open string spectrum on $\mathbb{C}^3
/(\mathbb{Z}_2  \times \mathbb{Z}_2)$: the matter }

\begin{figure}[ht]
\begin{center}
\includegraphics[width=7truecm]{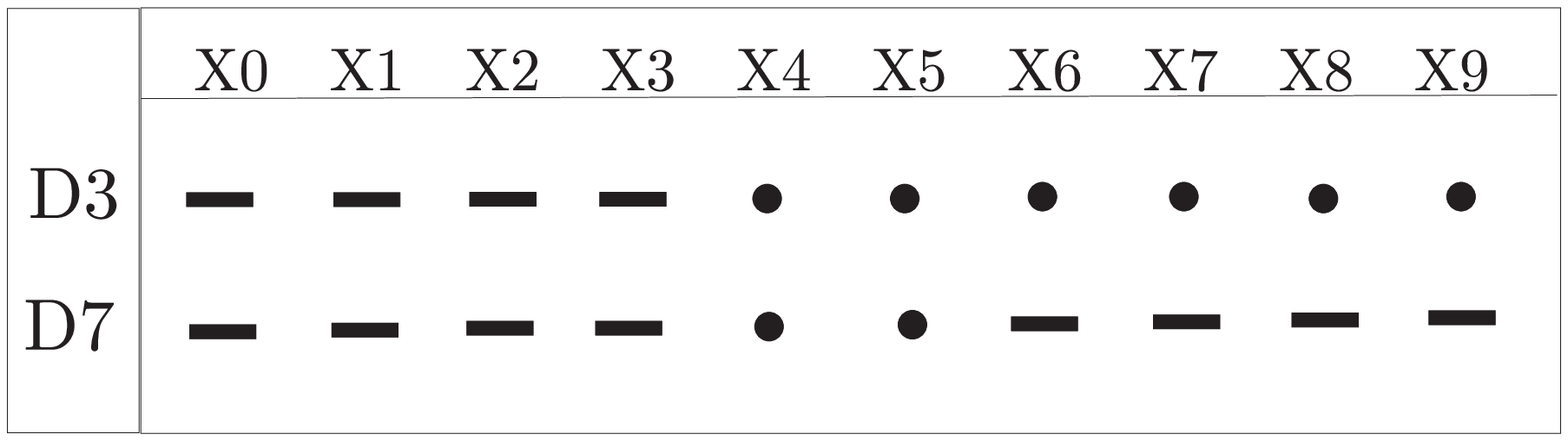}
\end{center}
\caption{Configuration of the D3-D7 branes system} \label{d3d7}
\end{figure}

In this subsection we introduce D7-branes in order to add extra
chiral matter to the gauge theory living on the stack of the
D3-branes discussed previously. The matter is given by the open
strings with one end attached to the D3-branes and the other one
to the D7-brane. Physical states associated to these strings
transform under the fundamental representation of the gauge group,
thus providing chiral matter fields while the number of D7-branes
can be regarded as a flavor index for them. In particular, we will
consider configurations in which the D7-branes extend in the
directions $x^{0}, \dots, x^{3}, x^{6}, \dots, x^{9},$ i.e.
partially along the orbifold, while the D3-branes remain in the
directions $x^{0}, \dots x^{3}$, being localized completely in the
D7-brane world-volume. The configuration is represented in
{}Fig.~{\ref{d3d7}}.
Also in this case we are interested in studying the bosonic
spectrum in the NS sector and the fermionic one in the R sector.
Supersymmetry allows one to deduce one sector from the other. Let
us therefore consider only the NS one. In this case the zero-point
energy is zero. {}Furthermore the open strings can have four
different boundary conditions along the directions $x^{\alpha}$
with $\alpha=6,7,8,9$ and therefore they have zero modes along
these directions that generate $2^{4/2}=4$ degenerate ground
states, which we label by their ``spins" in the $z^2$ and $z^3$
planes:
\begin{eqnarray}
\lambda_{37} | s_2 , s_3 \rangle, \
\end{eqnarray}
with  $\lambda_{37}$ being a Chan-Paton factor and
$\displaystyle{s_2=s_3=\pm 1/2}$, as implied by the GSO
projection. It turns out that a consistent definition of the
orbifold action on the physical states  is
\cite{Diaconescu:1999dt}:
\begin{eqnarray}
\left[ \gamma^{(D3)}_{h_i} \right] \lambda_{37}
\left[ {\gamma^{(D7)}_{h_i}}\right]^{-1} S_{h_i}(s_2,s_3) |s_2, s_3 \rangle =
\lambda_{37}' |s_2,s_3 \rangle
\end{eqnarray}
with $S_{h_i}(s_2,s_3)=e^{i\pi (s_2b^2_{h_i}+s_3b^3_{h_i})}$ and
\begin{eqnarray}
\gamma^{(D7)}_{e}&=&{\rm diag}(\mathbf{1}_{M_1,M_1},\,\,\,\,\mathbf{1}_{M_2,M_2},\,\,\,\,
\mathbf{1}_{M_3,M_3},\,\,\,\,\mathbf{1}_{M_4,M_4}) , \nonumber \\
\gamma^{(D7)}_{h_1}&=& {\rm diag}(\mathbf{1}_{M_1,M_1},\,\,\,\,\mathbf{1}_{M_2,M_2},
-\mathbf{1}_{M_3,M_3},-\mathbf{1}_{M_4,M_4}), \nonumber \\
\gamma^{(D7)}_{h_2}&=&e^{-i\frac{\pi}{2}}{\rm diag}(\mathbf{1}_{M_1,M_1},
-\mathbf{1}_{M_2, M_2}, \mathbf{1}_{M_3,M_3},-\mathbf{1}_{M_4,M_4}) ,
\nonumber \\
\gamma^{(D7)}_{h_3}&=&-e^{-i\frac{\pi}{2}}{\rm diag}
(\mathbf{1}_{M_1,M_1},-\mathbf{1}_{M_2,M_2},-\mathbf{1}_{M_3,M_3},
\mathbf{1}_{M_4,M_4}), \label{gamma7}
\end{eqnarray}
where $M_I$ represents the number of fractional $D7$ branes of
type $I$. The signs and the phase factors introduced in
(\ref{gamma7}) are consistent with group properties of
$\mathbb{Z}_2 \times \mathbb{Z}_2$ i.e. $h_1 \cdot h_2 = h_3$ and
$h_{i}^2=1$.

In the same way we can study the open string spectrum of the D7-D3 open
strings. In this case the states in the NS sector left invariant by the
orbifold projection are the ones satisfying:
 \begin{equation}
\left[ \gamma^{(D7)}_{h_i} \right] \lambda_{73}
\left[ \gamma^{(D3)}_{h_i} \right]^{-1} S_{h_i}(s_2,s_3)=
\lambda_{73}  .
\end{equation}
The physical states are those invariant under the orbifold action.
The physical spectrum (containing both bosons and fermions coming
from the R sector) is:
\begin{eqnarray}
\chi \equiv \Phi_{37} + \Phi_{73}  & \sim &(N_1,\overline{M}_3) +
(N_2, \overline{M}_4) + (N_3,\overline{M}_1) +
(N_4,\overline{M}_2)
\nonumber \\
&& +
(M_4,\overline{N}_1)
+(M_3,\overline{N}_2)+(M_2,\overline{N}_3)+(M_1,\overline{N}_4) .
\end{eqnarray}
The whole spectrum including both D3-D7 and D7-D3 open strings
contains eight chiral superfields as shown in Table \ref{tavola2} and, in general, is not gauge anomaly free.
\begin{table}
\begin{eqnarray*}
\begin{array}{cccc|cccc}
&D3&  {\rm Gauge-Groups}& & &D7&{\rm Global-Symmetries}& \\ \hline
\\
   U(N_1)  & U(N_2) & U(N_3) &  U(N_4)& U(M_1) & U(M_2) & U(M_3) & U(M_4) \\
  \hline \hline \\
  \square & 1& 1& 1 &1&1&\overline{\square}&1\\
  \overline{\square}& 1 & 1 & 1 &1&1&1&\square\\
  1&{\square}& 1&1   &1&1&1&\overline{\square}\\
 1&\overline{\square}&1&1   &1&1&\square&1 \\
 1&1&{\square}&1& \overline{\square}&1&1&1 \\
 1&1&\overline{\square}&1  &1&\square&1&1 \\
  1&1&1&{\square}&1&\overline{\square}&1&1 \\
 1&1&1&\overline{\square}&\square&1&1&1 \\
\end{array}
\end{eqnarray*}
\caption{Spectrum for the $D3$-$D7$ and $D7$-$D3$
systems.}\label{tavola2}
\end{table}
 However a simple consistent vector-like theory is
obtained for $M_1=M_2$ and $M_3=M_4$. This spectrum yields the
following contributions to the $\beta$-functions of the various
gauge groups:
\begin{eqnarray}
&& \Delta \beta (g_1) = \Delta\beta (g_2) = \frac{g^3}{16 \pi^2}
\left[ \frac{1}{2} (M_3 + M_4) \right], \\
&& \Delta\beta (g_3) = \Delta\beta (g_4) = \frac{g^3}{16 \pi^2}
\left[ \frac{1}{2} (M_1 + M_2) \right] \ .
\end{eqnarray}
{}Finally in ${\cal N}=1$ supersymmetry the interaction between
the different chiral fields can be encoded in the following cubic
superpotential \cite{AIQU} schematically written as:
\begin{eqnarray}
W= {\mbox Tr} \left( \Phi_1 [\Phi_2, \Phi_3 ] \right) +
\sum_{i=1}^{3} {\mbox Tr} \left( \Phi^{i} \, \chi^2 \right).
\end{eqnarray}

It is important to observe that if we turn off three of the gauge
groups by taking $N_2=N_3=N_4=0$ and $M_1=M_2=0$ while keeping
$N_1=N$ and $M_3=M_4=M$ non zero the resulting gauge theory is
exactly ${\cal N}=1$ super QCD with matter.

\section{The Fractional D3/D7 Bound State}
\subsection{The boundary state approach}

In this subsection we analyze a D3/D7 fractional branes system
located at the fixed point of the orbifold ${\rm I\!R}^{1,3}
\times \mathbb{C}^3/\left( \mathbb{Z}_2 \times \mathbb{Z}_2
\right)$ by using the boundary state approach.

The boundary state description for a fractional Dp-brane with a
given number of directions parallel to the orbifold
$\mathbb{C}^3/\left( \mathbb{Z}_2 \times \mathbb{Z}_2 \right)$ is
encoded in the following one-loop vacuum energy $Z$ of the open
strings stretched between two such Dp-branes:
\begin{equation}
Z= \int_{0}^{\infty} \frac{d s}{s} {\rm Tr}_{\rm  NS-R} \left[ P_{
\rm GSO} \left( \frac{\mathbf{1} + h_1+h_2+h_3}{4}\right) {\rm
e}^{-2\pi s (L_0-a) } \right] \ ,\label{Z}
\end{equation}
where $a=0 [1/2]$ in the R [NS] sector and the orbifold group
elements $(\mathbf{1}, \cdots, h_3)$ act both on the oscillators
and on the Chan-Paton factors of the open strings.

Alternatively, the interaction between two fractional Dp-branes
can be more easily studied by performing in eq.~(\ref{Z}) the
modular transformation $s\rightarrow 1/s$, that leads to the
closed string channel, and by rewriting $Z$ as a matrix element
between two boundary states $| Bp\rangle$:
\begin{eqnarray}
Z=\langle Bp| {\cal D} | Bp \rangle \ ,
\end{eqnarray}
with ${\cal D}$ being the closed string propagator. Our boundary
states are the sum of four terms, one untwisted obtained by
performing the modular transformation on the first term in eq.
(\ref{Z}), i.e. the one associated to the identity of the discrete
group, and three twisted sectors associated to the three non
trivial group elements $h_i$.

The sign in front of each twisted sector is determined by eqs.
(\ref{action}) and therefore we have four boundary states
describing the four different fractional branes that we can have
in this orbifold.  They are given by:
\begin{eqnarray}
&&| Bp>_1 =| Bp>_u+ | Bp>_{t_1}+ | Bp>_{t_2} +| Bp>_{t_3} \ , \nonumber\\
&&| Bp>_2 =| Bp>_u+ | Bp>_{t_1}- | Bp>_{t_2} -| Bp>_{t_3} \ ,\nonumber \\
&&| Bp>_3 =| Bp>_u- | Bp>_{t_1}+ | Bp>_{t_2} -| Bp>_{t_3} \ ,\nonumber \\
&&| Bp>_4 =| Bp>_u- | Bp>_{t_1}- | Bp>_{t_2} +| Bp>_{t_3} \ ,\label{bs}\\
\nonumber
\end{eqnarray}
where $p$ is the integer relative to the $Dp$-brane and $u$ labels
the untwisted sector while $t_i$ with $i=1,\ldots,3$ label the
twisted states.

The untwisted part of the boundary state, as already noticed in
Ref.~\cite{BDFM}, has the same structure as in flat space but with
an extra 1/2 factor due to the 1/4 factor of the orbifold
projector appearing in eq.~(\ref{Z}). The other terms in eqs.
(\ref{bs}), instead, apart an extra factor $1/\sqrt{2}$ due again
to the orbifold projector, are the same as those for the orbifold
$\mathbb{C}^2/\mathbb{Z}_2$.

Let us consider for instance the part in eq. (\ref{Z}) containing
just the $h_1$ group element. Since $h_1$ is the generator of
$\mathbb{Z}_2$ acting as a reflection along the directions $z_2$
and $z_3$, the associated twisted boundary states are the ones
corresponding to the twisted sector of the $\mathbb{Z}_2$ orbifold
acting on these directions. Similar arguments can be repeated for
the other two elements $h_2$ and $h_3$ acting respectively on
($z_1$, $z_3$) and on ($z_1$, $z_2$).

{}Finally for the orbifold $\mathbb{C}^3/(\mathbb{Z}_2 \times
\mathbb{Z}_2)$ we can directly use the twisted boundary state
given in Ref.~\cite{Bertolini:2001qa} for the orbifold
$\mathbb{C}^2/\mathbb{Z}_2$ and multiply it for an additional
factor $1/\sqrt{2}$.

Using the same logic we derive the couplings between the boundary
states and the closed string states. These couplings are
$1/\sqrt{2}$ of the corresponding ones for the orbifolds
$\mathbb{C}^2/\mathbb{Z}_2$ while we now have three twisted
sectors. {}Following Ref.~\cite{BDFM} we deduce the couplings of
the fractional D3-branes to the fluctuation of the $h_{MN}$, of
the untwisted four-form potential $C_4$ and of the twisted fields
${\widetilde b^i}$ and $A_{0123}^i$ \footnote{The couplings
written in the paper \cite{BDFM} are recovered when using the
relation between $\kappa$ and $\kappa_{\rm orb}$.}:
\begin{equation}
~_{u}\braket{B3}{h} = -\frac{T_3 }{2\kappa_{\rm orb}}\,V_4\,\,
h_{\mu}^{\,\,\, \mu} ~~~,~~~_{u}\braket{ B3}{ C_{4}} =
\frac{T_3}{2\,\kappa_{\rm orb}} \,V_4\,C_{0123} \label{untw9}
\end{equation}
\begin{equation}
~_{t_i}\braket{B3}{\widetilde b^i} = -\frac{T_3}{2 \,\kappa_{\rm
orb}} \,\frac{V_4}{2\pi^2\alpha'}\,\widetilde b^i ~~~,~~~_{t_i}
\braket{B3}{A_{4}^i} = \frac{T_3}{2 \,\kappa_{\rm
orb}}\,\frac{V_4}{2\pi^2 \alpha'}\,A_{0123}^i \label{twi86}
\end{equation}
where for a general $p$ $T_p = \sqrt{\pi}/(2\pi
\sqrt{\alpha'})^{p-3}~$,
 $\kappa_{\rm orb}= 2\, \kappa= 16 \pi^{7/2}(\alpha')^{2} g_{s}$,
$V_4$ is the (infinite) world-volume of the D3-brane and we
defined $b^i = b_0 + \widetilde{b}^{i}$, being $b_0$ the
background value of the $B_2$-flux \cite{Aspinwall:1995zi}:
\[
\int_{ {\cal C}_i } B_2 = 4 \pi^2 \alpha' \frac{1}{2} \equiv b_0
\]
for any $i$. {}From the above couplings one can write the boundary
action for a fractional D3-brane of type $I$ on this orbifold:
\begin{eqnarray}
&&S_{b,I}^{\rm D3} = - \frac{T_3}{2 \kappa_{\rm orb} } \int d^{4}
x \sqrt{- \det G_{\mu\nu}} \left[1 + \frac{1}{2 \pi^2 \alpha'}
\sum_{i=1}^{3} {\cal F}_I^i\, {\widetilde{b}}^i \right]
\nonumber\\
&& + \frac{T_3}{2\kappa_{\rm orb} } \int \left[C_4 \left(1 +
\frac{1}{2 \pi^2 \alpha'} \sum_{i=1}^{3}{\cal F}_I^i\,
{\widetilde{b}}^i \right) + \frac{1}{2 \pi^2 \alpha'}
\sum_{i=1}^{3}{\cal F}_I^i\, A_{4}^{i} \right] . \label{biac56}
\end{eqnarray}
The quantities ${\cal F}_I^i$ take in account the different signs appearing
in the definition of the boundary states (\ref{bs}) and their explicit form
is:
\begin{align}
&\vec{{\cal F}}_1=(+1,\,+1,\,+1)  & \quad  \vec{{\cal
F}}_2=(+1,\,-1,\,-1)& \quad
 \vec{{\cal F}}_3=(-1,\,+1,\,-1)  & \quad  \vec{{\cal F}}_4=(-1,\,-1,\,+1) .
\nonumber\\
\label{F}
\end{align}
In the case of a D7-brane, which is transverse only to the
direction $z_1$ and therefore partially extended along the
orbifold, we have to take into account the condition $M_1=M_2$ and
$M_3=M_4$, that as previously discussed, leads to an anomaly-free
gauge theory. The role played by this condition is to cancel, in
the boundary state describing this compound system, the
contributions coming from the twisted sectors $t_2$ and $t_3$ in
eq. (\ref{bs}). Hence we consider only the couplings between the
bulk fields and the twisted boundary state associated to the
element $h_1$ of the orbifold group.

By computing the couplings between the D7-fractional branes and
the bulk fields we get for the untwisted sector the following
expressions:
\begin{eqnarray}
&&~_{u}\braket{B7}{h} = -\frac{T_7}{2\kappa_{\rm orb}}\,\, V_8
{h_{\alpha}^{\,\,\mu}},\qquad  ~_{u} \braket{ B7}{ C_{8}} =
\frac{T_7}{2\,\kappa_{\rm orb}} V_8
\,C_{0,1,2,3,6,7,8,9}\nonumber\\
&&~_{u}\braket{ B7}{ \phi}=\frac{T_7}{2\, \kappa_{\rm orb}} \,V_8 \phi
 ,
\label{d7cu}
\end{eqnarray}
where $\phi$ is the dilaton and $C_8$ is the Ramond-Ramond 8-form
potential, while for the twisted fields we get:
\begin{equation}
~_{t_1}\braket{B7}{\widetilde b^1} = -\frac{T_3}{2 \,\kappa_{\rm
orb}} \,\frac{V_4}{ 8\pi^2\alpha'\, }\,\widetilde b^1
~~~,~~~_{t_1} \braket{B7}{A_{4}^1} = \frac{T_3}{2 \,\kappa_{\rm
orb}}\, \frac{V_4}{8\pi^2 \alpha' }\,A_{0123}^1 . \label{d7ct}
\end{equation}
Using the couplings (\ref{d7cu}) and (\ref{d7ct}) one can infer
the form of the linearized world-volume action for a D7 of type
$I$:
\begin{eqnarray}
&&S_{b,I}^{\rm D7} = - \frac{T_7}{2 \,\kappa_{ \rm orb}} \int
d^{8} x e^{\phi} \sqrt{- \det G_{\rho \sigma}} +\frac{T_7}{2
\,\kappa_{ \rm orb}}\int C_8
 +\nonumber\\
&& -\frac{T_3}{2\,\kappa_{\rm orb} }\frac{1}{8 \pi^2 \alpha' }
\int d^4 x \sqrt{-\det G_{\rho \sigma}} {\cal F}_I^1\,
{\widetilde{b} }^1  +\frac{T_3}{2\,\kappa_{\rm orb}} \frac{1}{8
\pi^2 \alpha' }\int {\cal F}_I^1\, A_{4}^{1}  +\cdots \label{d7b}
\end{eqnarray}
where $\rho, \sigma$ run over $0, \dots, 3, 6 \dots 9$. Here we
have enforced reparametrization invariance and the ellipses stay
for terms of higher order in $g_s$ which are not accounted for by
the boundary state approach.

{}Equation (\ref{d7b}) tells us that the boundary action for a
D7-brane of type 1 is coincident with the one for a D7-brane of
type 2; the same happens for type 3 and type 4. This means that,
due to the anomaly free gauge condition, we have only two
different kinds of fractional branes.

We now study the bound state made of $\mbox{N}_1$, $\mbox{N}_2$,
$\mbox{N}_3$  and $\mbox{N}_4$ D3-fractional branes respectively
of type $ 1,\ldots, 4$ extended along the direction $x^0,\ldots,
x^3$, and $M_1 = M_2$, $M_3 = M_4$ D7-fractional branes
respectively of type $1, \ldots, 4$ (with $1,2$ and $3,4$
identified) parallel to the D3-fractional branes and extended
along the directions $x^6,\ldots, x^9$  of the orbifold.

As explained in Ref.~\cite{{DiVecchia:1999rh},DiVecchia:1999fx},
the boundary state formalism allows to compute the asymptotic
behavior for large distances of the various fields in the
classical brane solution. In our case the boundary states of the
D3/D7  system are just linear combinations of the ones describing
the constituent branes. It follows that the asymptotic behavior of
the classical solution is the sum of the behaviors generated by
the single branes. We find that, at the leading order in $g_s$,
the dilaton and the R-R potentials are:
\begin{eqnarray}
\phi &\simeq&  \frac{ f_0(M) g_s}{2\pi}\log \rho_1 +\dots \
,\qquad  C_4 \simeq -\frac{Q_3}{r^4}~dx^0\wedge \cdots\wedge
dx^3+\dots \nonumber \\C_8 &\simeq&
\frac{f_0(M)g_s}{2\pi}\log\rho_1 ~dx^0\wedge \cdots\wedge
dx^3\wedge dx^6\cdots\wedge dx^9+\dots
 \label{dRR}
\end{eqnarray}
where $\rho_i=\sqrt{z_i\bar{z}_i}~/\epsilon$ for $i=1,2,3$ and
$\epsilon$ is  a regulator. The metric is:
\begin{equation}
ds^2\simeq H^{-1/2} \eta_{\mu\nu}
dx^{\mu} dx^{\nu} +
H^{1/2} \left( e^{- \phi} \delta_{ab}
 dx^adx^b +  \delta_{\alpha\beta} dx^{\alpha} dx^{\beta} \right)+\dots
\label{met}
\end{equation}
where
\begin{eqnarray} H = \left(1+ \frac{Q_3}{r^4} + \dots
\right),  \quad r=\sqrt{(x^4)^2+\cdots+(x^9)^2}, \quad  Q_3=
4\,\pi\, (\alpha')^2 g_s f_0(N)\ ,\end{eqnarray} with $\mu,
\nu=0,\cdots3$, $a,b=4,5$ and $\alpha,\beta=6,\cdots 9$. {}Finally
the asymptotic behavior of the twisted fields is given by (no sum
over $i$):
\begin{eqnarray}
 {\widetilde b}^i \simeq 4 \pi g_s \alpha' \left( f_i(N)
+\frac{f_1(M)}{4} \delta^i_1 \right) \log \rho_i\ , \quad  A^i
\simeq 4 \pi g_s \alpha'\left( f_i(N) +\frac{f_1(M)}{4}\delta^i_1
\right) \log \rho_i\ .\label{tbf}
\end{eqnarray}
In these formulas the factors $f_i(V)~(V=M,N)$ take into account
the different signs in front of the twisted boundary states
(\ref{bs}) and their explicit form is:
\begin{eqnarray}
f_0(V)=V_1+V_2+V_3 +V_4, && f_1(V)=  V_1 + V_2 - V_3-V_4, \nonumber\\
f_2(V)=  V_1 - V_2 + V_3  -V_4, && f_3(V)= V_1- V_2 - V_3+ V_4 .
\label{coupling}
\end{eqnarray}
In order to extend the previous analysis to all orders we should
solve the complete equation of motion of Type IIB supergravity in
this background, having as a source term the actions described by
eqs. (\ref{biac56}) and (\ref{d7b}). This requires a suitable
{\it ansatz} for the supergravity fields. The asymptotic ansatz,
given by eqs. (\ref{dRR})-(\ref{tbf}), is not the one that can
provide the complete solution in the case of branes localized
inside branes, which is exactly the configuration we are
considering here \cite{Itzhaki:1998uz}.

In the following we will be interested in the classical profile of
the RR-form of lower degree and therefore we rewrite the previous
solution in terms of the axion $C_0$ and the twisted scalar $c$
related respectively to the Hodge duals of the RR-potentials $C_8$
and $A_4^i$. At first order in $g_s$ we have:
\begin{equation}
^*dC_8=-d C_0 \ ,\qquad ^{*6}dc^{i}=dA_4^i \ , \label{ccdual}
\end{equation}
whose explicit solutions from eqs. (\ref{dRR}) and (\ref{tbf})
are:
\begin{eqnarray}
C_0 \simeq\frac{f_0(N)}{2 \pi} g_s \theta_1\ ,\qquad c^{i}\simeq -
4 \pi g_s \alpha' \left(f_i(N) +\frac{f_1(M)}{4}\delta_1^i\right)
\theta_i \label{ccdual1}
\end{eqnarray}
where $\theta_i= {\rm tan }^{-1}x^{2i+3}/x^{2i+2}$ for $i=1,2,3$.
We combine together the dilaton, the axion and the twisted scalars
$b$ and $c$ in two complex quantities relevant for the gravity
dual:
\begin{eqnarray}
&&\tau~~~\simeq C_0 +ie^{-\phi}= i-\frac{g_s}{2\pi}f_0(M)\log w_1,
\label{dil}\\
&& \gamma^{(1)}\simeq\tau b^{1}+c^{1}=  i 2\pi^2\alpha'+ 4\pi g_s
\alpha' i \left( f_1(N) + \frac{(f_1(M)-f_0(M) }{4}\right) \log
w_1,
\label{gamma1}\\
&& \gamma^{(2)}\simeq\tau b^{2}+c^{2}=  i2\pi^2\alpha'+ 4\pi g_s
\alpha' i \left( f_2(N) \log w_2-\frac{f_0(M) }{4} \log
w_1\right),
\label{gamma2}\\
&& \gamma^{(3)}\simeq\tau b^{3}+c^{3}=  i2\pi^2\alpha'+ 4\pi g_s
\alpha' i \left( f_3(N) \log w_3 -\frac{f_0(M) }{4} \log w_1
\right), \label{gamma3}
\end{eqnarray}
where we have defined $w_i=z_i/\epsilon$ .

We see from eqs. (\ref{dil})-(\ref{gamma3}) that the dilaton  and
the $\gamma$'s are holomorphic functions of the $w_i$'s, at least
at the first order in $g_s$. This is  a property common to all
classical solutions having fractional branes as a source and it is
a consequence of supersymmetry \cite{Grana:2001xn}.

\subsection{Stability of the system and massless tadpole cancellation}
We now investigate the stability of the system related to the
one-loop massless tadpole cancellation. The generic boundary state
describing our D3/D7 fractional system is the following linear
combination of the boundary states introduced in the previous
paragraph:
\begin{equation}
|B3/7\rangle_{I,J} = |B3\rangle_I \, +|B7\rangle_J .
\end{equation}
Two D-branes interact at the tree level via the exchange of closed
strings and the interaction is:
\begin{equation}
\frac{\alpha' \pi }{2} \int_{0}^{\infty} d\,t ~_{I,J} \langle
B3/7| e^{-\pi t\left(L_0+\bar{L}_0-2\,a\right)} |B3/7\rangle_{I,J}
\label{RR2}
\end{equation}
where $a=0$ in the twisted sectors. Due to the BPS no-force
condition even before explicitly evaluating the previous
expression we expect a vanishing result. This is consistent with
our bound state which preserves ${\cal N}=1$ and with the massless
tadpole cancellation requirement.

By separating the various contributions in the correlator
(\ref{RR2}) and considering  only the twisted sectors encoding the
orbifold information we obtain:
\begin{eqnarray}
&&\frac{\alpha' \pi }{2} \int_{0}^{\infty} d\,t ~_{t_i}\langle Bp|
e^{-\pi t\left(L_0+\bar{L}_0-2\,a\right) }    |Bp\rangle_{t_i} = \nonumber\\
&&=\frac{\alpha^{\prime}\pi}{4} \frac{T_{r_i}^2 V_{r_i+1} }{
2^{s_i} (2\pi^2 \alpha')^{\frac{9-r_i}{2} }} \int_0^\infty
\frac{d\, t}{t^{\frac{5-r_i}{2}}} \prod_{n=1}^{\infty}
  \frac{ \left(1+e^{-2n\pi t}\right)^4 \left(1+e^{-(2n-1)\pi t}\right)^4}
{ \left(1-e^{-2n\pi t}\right)^4 \left(1-e^{-(2n-1)\pi t}\right)^4}
\left(1|_{\rm NS-NS}-1|_{\rm RR}\right),
\nonumber\\
&&\label{56}\\
&&\frac{\alpha' \pi }{2} \int_0^{\infty} \frac{d\,t}{t}
~_{t_i}\langle B3|
e^{-\pi t\left(L_0+\bar{L}_0-2\,a\right)}    |B7\rangle_{t_i}=\nonumber\\
&&= \frac{\alpha^{\prime}\pi}{16} \delta_{{t_i},{t_1}}
\frac{T_3^2V_4 }{(2\pi^2 \alpha')^3} \int_0^\infty \frac{d\, t}{t}
\prod_{n=1}^{\infty}
  \frac{ \left(1+e^{-2n\pi t}\right)^4 \left(1-e^{-(2n-1)\pi t}\right)^4}
{\left(1-e^{-2n\pi t}\right)^4 \left(1+e^{-(2n-1)\pi t}\right)^4}
\left(1|_{\rm NS-NS}-1|_{\rm RR}\right),\nonumber\\
&&  \label{57}
\end{eqnarray}
where $r_i$'s are the world-volume directions outside the sector
$i$ of the orbifold and $s_i$'s are the world-volume directions
inside the sector $i$ of the orbifold. By sector $i$ of the
orbifold we mean the two complex coordinates non trivially
transformed by the group element $h_i$ and hence we have for the
D3-brane $r_i=3$ ($s_i=0$) for $i=1,2,3$ while for the D7-brane
$r_1=3$, $r_2=r_3=5$ ($s_1=4$, $s_2=s_3=2$). Equations(\ref{56})
and (\ref{57}) show the cancellation between the contributions due
respectively to the NS-NS and the RR sectors. Each of the two term
corresponds to a one-loop massless tadpole. For such a term, we
 see that in the limit $t\rightarrow \infty$ we have only
logarithmic divergences for the contributions coming from
sandwiching the closed string propagator with the $D3$ boundary
state since $r_i=3$. The same holds true when considering one
boundary state of type $D3$ and the other of type $D7$. In the
spirit of the open/closed sting duality these logarithmic
divergences are related to the UV divergences of the gauge theory.
When sandwiching the propagator between two $D7$ boundary states
we observe in the twisted sectors $t_2$ and $t_3$ the rise of
linear divergences which are dangerous since they lead to
supersymmetry and/or Poincar\'e breaking \cite{Kakushadze:2001mu}.

As can be seen from the states in eqs.~(\ref{bs}) this divergence
can be cured by imposing $M_1=M_2$ and $M_3=M_4$. Such a condition
yields, as expected, a consistent gauge theory free from gauge
anomalies as seen in section 2.

\section{Supergravity analysis of the dual gauge theory anomalies}

In this section we discuss in detail from the dual supergravity
point of view the trace and chiral anomalies of the gauge theory
supported by a bound state made of D3/D7 branes.

As mentioned above, in our case the gauge theory living on the D3-world volume
is an ${\cal N}=1$ super Yang-Mills theory with gauge
group $SU(N_1)\times SU(N_2)\times SU(N_3)\times SU(N_4)$. Here we
do not consider the effects of the abelian $U(1)$
factors\footnote{The effects are subleading in the large $N$
limit.}. The theory has twelve chiral multiplets in the
bifundamental representation of the gauge group due to the D3/D3
open strings and chiral matter multiplets, both in the fundamental
and in the antifundamental representation, due to the D3/D7
strings.

At the classical level this theory is invariant under an abelian
$\hat{U}{(1)}_R$ and the scale symmetry. We note that it is always
possible to define a new anomalous free $U(1)_R$ symmetry by
considering a suitable choice of the chiral super-multiplets. The
scalars of the gauge theory have canonical dimension one and
R-$\hat{U}{(1)}_R$ charge 2/3 as it follows from the cubic
superpotential, therefore under a combined chiral and scale
transformation with parameters respectively $\alpha$ and $\sigma$
they transform as:
\begin{equation}
\phi'= e^{ \sigma \, + \,i\, \frac{2}{3} \alpha } \phi \
.\label{sct}
\end{equation}
These global symmetries are broken at the quantum level by
anomalies, and the respective currents are given by:
\begin{eqnarray}
&&\partial_{\mu} D^{\mu}=\frac{\beta(g_{\rm YM})}{2\,g_{\rm YM}^3}
F_{\mu\nu}^a F^{a\,\mu\nu} , \\
&&\partial_{\mu}J_R^{\mu}=\sum_{\rm  fermions} \frac{Q_R\, C_2(T) }{16 \pi^2}
F_{\mu\nu}^a \tilde{F}^{a\,\mu\nu} ,
\label{anomalie}
\end{eqnarray}
where $Q_R$ is the R-charge of the fermions, $C_2(T)$ is the
quadratic Casimir in the representation $T$ of the gauge group and
$\tilde{F}^{a\,\mu\nu}=\frac{1}{2} \epsilon^{\mu\nu\rho\sigma}
F_{a\,\rho\sigma}$. {}For infinitesimal transformations these
equations imply \cite{Peskin:ev}:
\begin{equation}
\frac{1}{g_{\rm YM}^2}\rightarrow \frac{1}{g_{\rm YM}^2} - \frac{2
\beta(g)}{ g_{\rm YM}^3} \sigma\hspace{2cm} \theta_{\rm YM}
\rightarrow \theta_{\rm YM} -2\alpha \sum_{\rm fermions}Q_R C_2
(T) \ .\label{anomalie1}
\end{equation}
In a supersymmetric theory, such as the one we are considering,
these two anomalies are in the same  supermultiplet; it follows
that, in a renormalization scheme preserving the ABJ theorem, the
$\beta$-function appearing in the previous expression is the one
at the 1-loop level. Indeed in the present case, by using the
spectrum discussed in section (2), one obtains the 1-loop
$\beta$-function for each $U(N_I)$ leading to:
\begin{eqnarray}
&&\partial_{\mu} D^{\mu}_I=- \frac{1}{2 (4\pi)^2}\left[3\,N_I-
\sum_{J\neq I=1}^4N_J - \frac{1}{2} \left( M_{I+2} + M_{-I+5}
\right) \right]F_{(I)\,\mu\nu}^{a} F_{(I)}^{a\,\mu\nu} \ ,
\label{anomsc}
\end{eqnarray}
where $M_I$ satisfies the condition $M_{I+4}\equiv M_I$.

Analogously, since the R-charge for the gauginos is one while for
the chiral fermions it is -1/3, the chiral anomaly for each gauge
group turns out to be:
\begin{eqnarray}
&&\partial_{\mu}J_{R,(I)}^{\mu} = \frac{1 }{(4 \pi)^2}\left[N_I
-\frac{1}{3}\sum_{J\neq I=1}^4 N_J
-\frac{1}{6}\left(M_{I+2}+M_{-I+5}\right)\right] F_{(I)\,
\mu\nu}^a \tilde{F}_{(I)}^{a\,\mu\nu} \ . \label{anomchi}
\end{eqnarray}
By defining the well-known complex gauge coupling
\begin{equation}
\tau_{\rm YM}=\frac{4\,\pi}{g_{\rm YM}^2}i +\frac{\theta_{\rm YM}
}{2\pi} \ ,\end{equation} and by using the expression of the one-loop
beta functions and chiral anomaly together with
eq.~(\ref{anomalie1}), one can easily see that $\tau_{\rm YM}$ transforms as:
\begin{eqnarray}
&&\tau_{\rm YM}^I \rightarrow \tau_{\rm YM}^I + \frac{i}{2\pi}
\left[3\,N_I- \sum_{J\neq I=1}^4N_J - \frac{1}{2} \left( M_{I+2} +
M_{-I+5} \right) \right]\left(\sigma +\frac{2}{3} i\alpha\right) \
.
\nonumber\\
&&\label{taug}
\end{eqnarray}
Let us now analyze how the scale and chiral anomalies are realized
in supergravity and to this aim let us consider the
Dirac-Born-Infeld action and the Wess-Zumino term for a stack of
$N_I$ fractional D3-branes given by eq. (\ref{biac56}).
Turning on a gauge field on the world-volume of the  branes and
expanding the boundary action in the supergravity background up to
quadratic terms in the derivatives one gets:
\begin{eqnarray}
&& S_{\rm gauge}^I=-\frac{1}{16
\pi g_s}\int d^4x \sqrt{-{\rm det} G}\,
 e^{-\phi} G^{\mu\rho}G^{\nu\sigma}\,\frac{1}{4} F_{\mu\nu}^a
F^{a}_{\rho\sigma} \left[1 +
\frac{1}{2\pi^2\alpha'}\sum_{i=1}^3{\cal F}_I^i\,
\tilde{b}^{i}\right]
\nonumber\\
&& +\frac{1}{64 \pi g_s}\int d^4x \left[C_0\left(
1+\frac{1}{2\pi^2\alpha'} \sum_{i=1}^3{\cal F}_I^i\, \tilde{b}^{i}
\right)+ \frac{1}{2\pi^2\alpha'}\sum_{i=1}^3{\cal F}_I^i
c^{i}\right] F_{\mu\nu}^a {\tilde F}^{a\,\mu\nu} \ ,
\end{eqnarray}
where for simplicity we dropped the index $I$ of the gauge fields
and the metric $G$ is the pull-back to the brane world-volume. The
previous boundary action, when evaluated in the supergravity
background obtained in the previous section, yields the (gauge
part of) ${\cal N}=1$ super YM theory with gauge group $SU(N_I)$.
More explicitly one obtains:
\begin{equation}
S_{\rm gauge}^I= - \frac{1}{g_{\rm YM}^2}\int d^4 x \frac{1}{4}
F_{\mu\nu}^a F^{a\,\mu\nu}+\frac{\theta_{\rm YM}}{32\pi^2} \int
d^4x F_{\mu\nu}^a \tilde{F}^{a\,\mu\nu} \ , \label{sgauge}
\end{equation}
where the indices are raised by the flat metric $\eta_{\mu\nu}$ and
\begin{equation}
 \frac{1}{g_{\rm YM}^2}= \frac{1}{16\pi g_s} e^{-\phi}
\left[1 + \frac{1}{2\pi^2\alpha'}\sum_{i=1}^3{\cal F}_I^i\,
\tilde{b}^{i}\right] \ ,
\end{equation}
\begin{equation}
 \theta_{\rm YM}=\frac{\pi}{2\, g_s}\left[C_0\left( 1+\frac{1}{2\pi^2\alpha'}
\sum_{i=1}^3 {\cal F}_I^i\, \tilde{b}^{i}\right) +
\frac{1}{2\pi^2\alpha'}\sum_{i=1}^3 {\cal F}_I^i\, c^{i}\right] \
.
\end{equation}
It is worthwhile to observe that in eq. (\ref{sgauge}) all the
dependence on the function $H$ of the metric disappears. We have
also to include matter in the fundamental and bifundamental
representations of the gauge groups. Then, by using eqs.
(\ref{dRR}) and (\ref{tbf}) the gauge coupling and the theta angle
turn out to be related to the bulk fields by the following
expressions:
\begin{equation}
\frac{1}{g_{\rm YM}^2}=\;\frac{1}{16\pi g_s}+ \frac{1}{8\, \pi^2}
\sum_{i=1}^3 \left[{\cal F}_I^i\,f_i(N) + \frac{ {\cal F}_I^1
\,f_1(M)-f_0(M)}{4}  \delta^i_1  \right] \log  \rho_i \label{gymg}
\ ,\end{equation}
\begin{equation}
 \theta_{\rm YM} = -\sum_{i=1}^3  \left[  {\cal F}_I^i \, f_i(N) +\frac{
{\cal F}_I^1 f_1(M) -f_0(M)}{4}\delta_1^i\right]\theta_i \ .
\label{tetaym}
\end{equation}
We note that the method considered above does not rely on the
probe technique. The ingredients we are using are the holographic
identification among the world-volume fields and the bulk
supergravity quantities. In this spirit one has to consider the
relations (\ref{gymg}) and (\ref{tetaym}) as directly generated by
the low-energy limit of the world-volume action of the fractional
branes.

Analogously with the gauge theory we combine the previous
equations and construct the supergravity realization of the
complex coupling $\tau_{\rm YM}$:
\begin{equation}
\tau_{\rm YM}^I= \frac{i}{4\,g_s} + \frac{i}{2\,\pi}
\sum_{i=1}^{3} {\cal F}_I^i f_i(N) \log w_i+
\frac{i}{2\,\pi}\frac{{\cal F}_I^1\,f_1(M)-f_0(M)}{4}\log w_1 \
.\label{tauym}
\end{equation}
In supergravity the coordinates $z_i$ are holographically
identified with the scalar components $\phi_i$ of the chiral
superfield $\Phi_i$. By using this identification we deduce that
under a chiral and scale transformation the coordinates transverse
to the brane transform according to the eq. (\ref{sct}), i.e.
$z_i\rightarrow e^{\sigma+\frac{2}{3}i\alpha} z_i$. The classical
profile of the twisted fields are not invariant under these
transformations, showing the  explicit breaking of the related
symmetries. In particular the complex combination given in eq.
(\ref{tauym}) behaves as:
\begin{equation}
\tau_{\rm YM}^I\rightarrow \tau_{\rm YM}^I+\frac{i}{2\,\pi} \left[
\sum_{i=1}^{3} {\cal F}_I^i f_i(N)  + \frac{{\cal
F}_I^1\,f_1(M)-f_0(M)}{4}\right] \left( \sigma\, + \frac{2}{3} i
\alpha\right) \ ,\label{tautr}
\end{equation}
where
\begin{eqnarray}
&& {\cal F}_1^1f_1(M)-f_0(M)={\cal F}_2^1f_1(M)-f_0(M)= - 2 (M_3\, +\, M_4)
\nonumber \\
&&{\cal F}_3^1f_1(M)-f_0(M)={\cal F}_4^1f_1(M)-f_0(M)= - 2 (M_1\, +\, M_2)
\nonumber \\
&&\sum_{i=1}^{3} {\cal F}_I^i f_i(N)=3 N_I -\sum_{J\neq I=1}^4N_J
\ .\label{exp}
\end{eqnarray}
Combining eq. (\ref{tautr}) with eq. (\ref{exp}) we recover  the right
$\beta$-function and chiral anomaly from the supergravity side.

\vspace{.5 cm}
In conclusion, by considering a suitable D3/D7 bound state living
on an orbifold preserving ${\cal N}=1$, as
$\mathbb{C}^3/(\mathbb{Z}_2 \times \mathbb{Z}_2)$, we obtained a
${\cal N}=1$ supersymmetric theory with matter fields. We have
studied the supergravity background determined by this brane
configuration by using the boundary state formalism and from this
classical solution we have got information about one-loop
quantities of the gauge theory. In so doing, we have shown how the
gauge/gravity correspondence may be used to analyze the quantum
aspects, at least at perturbative levels, of the dual string-like
standard model realized by considering branes at orbifold
singularities in the spirit of the bottom-up approach. It would be
nice to extend our analysis to the infrared regime of the gauge
theory by finding a complete non-singular solution. This analysis
would be performed by deforming, in an appropriate way, the
orbifold singularities and finding a suitable ansatz valid also
very near to the brane.

\vskip 1.cm \noindent {\large {\bf Acknowledgments}} \vskip 0.2cm
\noindent We thank Paolo Di Vecchia for careful reading of the
manuscript and useful discussions.
R.M. thanks L. Cappiello for discussions. R.M. and F.P. thank Nordita for
the kind hospitality in different stages of this work. The work of
F.S. is supported by the Marie--Curie fellowship under contract
MCFI-2001-00181.



\begin{thebibliography}{99}


\bibitem{AIQU} See for G. Aldazabal, L. E. Ibanez, F. Quevedo, A. M. Uranga,
{\it D-branes at singularities: A bottom-up approach to the string
embedding of the standard model}, hep-th/0005067 and references
therein.

\bibitem{Bertolini:2001gq}
M.~Bertolini, P.~Di Vecchia and R.~Marotta,
{\it N = 2 four-dimensional gauge theories from fractional branes},
hep-th/0112195.

\bibitem{Klebanov:1999rd}
I.~R.~Klebanov and N.~A.~Nekrasov,
{\it Gravity duals of fractional branes and logarithmic RG flow},
Nucl.\ Phys.\ B {\bf 574}, 263 (2000) hep-th/9911096.
\bibitem{Marotta:2002ns}
R.~Marotta and F.~Sannino, {\it N = 1 super Yang-Mills
renormalization schemes for fractional branes}, hep-th/0207163.
\bibitem{KS}
I. R. Klebanov and M. J. Strassler,
{\it Supergravity and a confining gauge theory: Duality cascades and $\chi$SB-Resolution of Naked Singularities},
JHEP {\bf 0008}, 052 (2000) hep-th/0007191.
\bibitem{MN}
J. M. Maldacena and C. Nu\~{n}ez,
{\it Towards the large N limit of pure {\cal N}=1 super Yang Mills},
Phys. Rev. Lett. {\bf 86} (2001) 588,
hep-th/0008001.



\bibitem{ABCPZ} R. Apreda, F. Bigazzi, A. L. Cotrone, M. Petrini,
A. Zaffaroni, {\it  Some Comments on N=1 Gauge Theories from
Wrapped Branes}, Phys.Lett. B{\bf 536} (2002)  16, hep-th/0112236.


\bibitem{DLM} P. Di Vecchia, A. Lerda and P. Merlatti,
{\it N=1 and N=2 super Yang-Mills from wrapped branes},
hep-th/0207039.
\bibitem{I} E. Imeroni,
{\it On the N=1 beta-function from the conifold},
hep-th/0205216.
\bibitem{OS} P. Olesen and F. Sannino,
{\it N=1 super Yang-Mills from supergravity: the UV-IR connection},
hep-th/0011077.
\bibitem{Bertolini:2000dk}
M.~Bertolini, P.~Di Vecchia, M.~Frau, A.~Lerda, R.~Marotta and
I.~Pesando,
{\it Fractional D-branes and their gauge duals},
JHEP {\bf 0102} 014 (2001), hep-th/0011077.
\bibitem{Polchinski:2000mx}
J.~Polchinski,
{\it N = 2 gauge-gravity duals},
Int.\ J.\ Mod.\ Phys.\ A {\bf 16} 707 (2001),  hep-th/0011193.
\bibitem{Frau:2000gk}
M.~Frau, A.~Liccardo and R.~Musto, {\it The geometry of fractional
branes}, Nucl.\ Phys.\ B {\bf 602} 39 (2001), hep-th/0012035.
\bibitem{Billo:2001vg}
M.~Bill\`o, L.~Gallot and A.~Liccardo,
{\it Classical geometry and gauge duals for fractional branes on
ALE  orbifolds},
Nucl.\ Phys.\ B {\bf 614} 254 (2001), hep-th/0105258.
\bibitem{Grana:2000jj}
M.~Grana and J.~Polchinski,
{\it Supersymmetric three-form flux perturbations on AdS(5)},
Phys.\ Rev.\ D {\bf 63} 026001 (2001), hep-th/0009211.
\bibitem{Bertolini:2001qa}
M.~Bertolini, P.~Di Vecchia, M.~Frau, A.~Lerda and R.~Marotta,
{\it N = 2 gauge theories on systems of fractional D3/D7 branes},
Nucl.\ Phys.\ B {\bf 621} 157 (2002), hep-th/0107057.
\bibitem{Johnson:1999qt}
C.~V.~Johnson, A.~W.~Peet and J.~Polchinski,
{\it Gauge theory and the excision of repulson singularities},
Phys.\ Rev.\ D {\bf 61} 086001 (2000), hep-th/9911161.
\bibitem{Diaconescu:1999dt}
D.~E.~Diaconescu and J.~Gomis,
{\it Fractional branes and boundary states in orbifold theories},
JHEP {\bf 0010} 001 (2000), hep-th/9906242.
\bibitem{BDFM} M. Bertolini, P. Di Vecchia, G. Ferretti and R. Marotta,
{\it Fractional Branes and {\cal N=1} Gauge Theories}
Nucl. Phys. B{\bf 630} (2002) 222 hep-th/0121817.
\bibitem{DiVecchia:1999rh}
P.~Di Vecchia and A.~Liccardo,
{\it D branes in string theory. I},
hep-th/9912161.
\bibitem{DiVecchia:1999fx}
P.~Di Vecchia and A.~Liccardo,
{\it D-branes in string theory. II},
hep-th/9912275.
\bibitem{BDFLM02} M. Bertolini, P. Di Vecchia, M. Frau, A. Lerda and R. Marotta, {\it More Anomalies from Fractional Branes}, Phys. Lett. B {\bf 540} (2002)
hep-th/0202195.
\bibitem{Klebanov:2002gr}
I.~R.~Klebanov, P.~Ouyang and E.~Witten,
{\it A gravity dual of the chiral anomaly},
Phys.\ Rev.\ D {\bf 65} 105007 (2002), hep-th/0202056.

\bibitem{Aspinwall:1995zi}
P.~S.~Aspinwall,
{\it Enhanced gauge symmetries and K3 surfaces},
Phys.\ Lett.\ B {\bf 357} 329 (1995), hep-th/9507012.
\bibitem{Itzhaki:1998uz}
N.~Itzhaki, A.~A.~Tseytlin and S.~Yankielowicz,
{\it Supergravity solutions for branes localized within branes},
Phys.\ Lett.\ B {\bf 432} 298 (1998), hep-th/9803103.

\bibitem{Grana:2001xn}
M.~Grana and J.~Polchinski,
{\it Gauge / gravity duals with holomorphic dilaton},
Phys.\ Rev.\ D {\bf 65} 126005 (2002), hep-th/0106014.
\bibitem{Kakushadze:2001mu}
Z.~Kakushadze and R.~Roiban,
{\it Brane-bulk duality and non-conformal gauge theories},
JHEP {\bf 0103} 043 (2001), hep-th/0102125.

\bibitem{Peskin:ev}
M.~E.~Peskin and D.~V.~Schroeder, {\it An Introduction To Quantum
Field Theory}, {\it  Reading, USA: Addison-Wesley (1995) 842 p}.




\end{thebibliography}
\end{document}